\documentclass{article}
\usepackage{spconf,amsmath,graphicx}

\usepackage{todonotes}

\usepackage{enumitem}
\setlist{nosep, leftmargin=14pt}

\usepackage{mwe} 

\title{Morphological Change Forecasting for Prostate Glands using Feature-based Registration and Kernel Density Extrapolation}

\makeatletter
\def\@name{\emph{Qianye Yang$^{1}$},\emph{ Tom Vercauteren$^{2}$},\emph{ Yunguan Fu$^{1,3}$},\emph{ Francesco Giganti$^{4,5}$},\emph{ Nooshin Ghavami$^{2}$}, \\ \emph{ Vasilis Stavrinides$^{4}$},\emph{ Caroline Moore$^{4}$},\emph{ Matt Clarkson$^{1}$},\emph{ Dean Barratt$^{1}$},\emph{ Yipeng Hu$^{1}$}}
\makeatother

\address{
$^{1}$ CMIC and WEISS, University College London, UK
\and
$^{2}$ School of Biomedical Engineering and Imaging Sciences, King’s College London, UK
\and
$^{3}$ InstaDeep, London, UK
\and
$^{4}$ Division of Surgery \& Interventional Science, University College London, London, UK 
\and 
$^{5}$ Department of Radiology, University College London Hospital NHS Foundation Trust, London, UK
}
\begin{document}
%
\maketitle
\begin{abstract}
Organ morphology is a key indicator for prostate disease diagnosis and prognosis. For instance, In longitudinal study of prostate cancer patients under active surveillance, the volume, boundary smoothness and their changes are closely monitored on time-series MR image data. In this paper, we describe a new framework for forecasting prostate morphological changes, as the ability to detect such changes earlier than what is currently possible may enable timely treatment or avoiding unnecessary confirmatory biopsies. In this work, an efficient feature-based MR image registration is first developed to align delineated prostate gland capsules to quantify the morphological changes using the inferred dense displacement fields (DDFs). We then propose to use kernel density estimation (KDE) of the probability density of the DDF-represented \textit{future morphology changes}, between current and future time points, before the future data become available. The KDE utilises a novel distance function that takes into account morphology, stage-of-progression and duration-of-change, which are considered factors in such subject-specific forecasting. We validate the proposed approach on image masks unseen to registration network training, without using any data acquired at the future target time points. The experiment results are presented on a longitudinal data set with 331 images from 73 patients, yielding an average Dice score of 0.865 on a holdout set, between the ground-truth and the image masks warped by the KDE-predicted-DDFs.
\end{abstract}
\begin{keywords}
Feature-based registration, kernel density estimation, longitudinal data, active surveillance, deep learning
\end{keywords}
\section{Introduction}
\label{sec:intro}
Radiological examination of the prostate gland, together with suspicious regions of pathological interest, has been an important tool in diagnosing and making prediction of prognosis for patients with prostate disease, such as differentiating between benign prostatic hyperplasia and clinically significant cancer. With emerging localised treatment options and multiparametric MR imaging proposed for screening early stage cancers, attention has been directed towards the morphological changes, both on the glands and the lesions, observed on time-series MR imaging data \cite{lovegrove2018prostate}. For example, the changes in size, smoothness and degree of conspicity, have been demonstrated to be associated with cancer progression or regression \cite{stavrinides2019mri}. Quantifying these changes, such as using medical image registration techniques \cite{yang2020longitudinal}, provides a means to model and better understand the disease progression process.

In the literature, image registration has been classified as intensity-based and feature-based approaches, depending on whether an explicit feature extraction step is required \cite{shen2002hammer}. The image features include points \cite{baum2020multimodality}, contours \cite{gu2013contour}, surfaces \cite{mazumder2020validnet}, anatomical structures \cite{hu2018weakly} or their combinations. In this report, we extend this terminological convention to recently-proposed registration algorithms based on deep learning, by defining feature-based registration networks for those take image features as input, as opposed to using intensity-valued images as network input \cite{balakrishnan2019voxelmorph, hu2018weakly}. The network input determines what is needed during inference, therefore the feature-based registration methods defined as such may take advantage of the currently well-established, better-validated automatic segmentation algorithms. Arguably, `transforming' images into a relatively image-acquisition-independent feature representation of morphology may improve generalisation of the subsequent morphological analysis to data acquired with different scanners or protocols. Therefore, in this work, we focus on testing a proposed forecasting approach using registration-quantified morphological changes, in which gland masks are used. However, we acknowledge that, in future studies, direct utilising or combining image intensities \cite{yang2020longitudinal,orczyk20173d} and other forms of features should be investigated.

Analysing longitudinal prostate MR images informs treatment intervention or alternative diagnostic assessment. Forecasting these clinical outcomes before the next imaging brings substantial patient benefits. For example, detecting cancer when they are small makes them more amenable to less-invasive focal therapy; and predicting the lack of clinical significance can avoid unnecessary needle biopsies which are costly and associated with risk of complications. This work investigates a kernel density estimation (KDE) method with a novel kernel distance function measuring not only the difference in morphology, as in most existing KDE-based morphological studies \cite{vuollo2016analyzing, hu2015population}, but also the differences in duration-of-change, stage-of-progression and subject-specificity. We show experimentally that the proposed KDE models are able to predict how the prostate gland transforms to a future target time at individual voxel positions. 

The contributions from this study are summarised as follows: 1) We develop and validate a feature-based registration network which quantifies the morphological changes between image masks, reporting its performance in both efficiency and accuracy; 2) We introduce a new kernel function for modelling complex longitudinal changes, such that the future morphological changes can be extrapolated using data acquired only at current time; and 3) Our forecasting approach is evaluated on a novel clinical application, forecasting the gland morphological changes for prostate cancer patients from an active surveillance program, with quantitative results on a holdout set of real clinical data.

\section{Method}
\label{sec:method}
\subsection{Feature-based registration for morphological change}
\label{sec:feature-based registration}
\begin{figure}[htb]
    \centering
    \includegraphics[width=8.5cm]{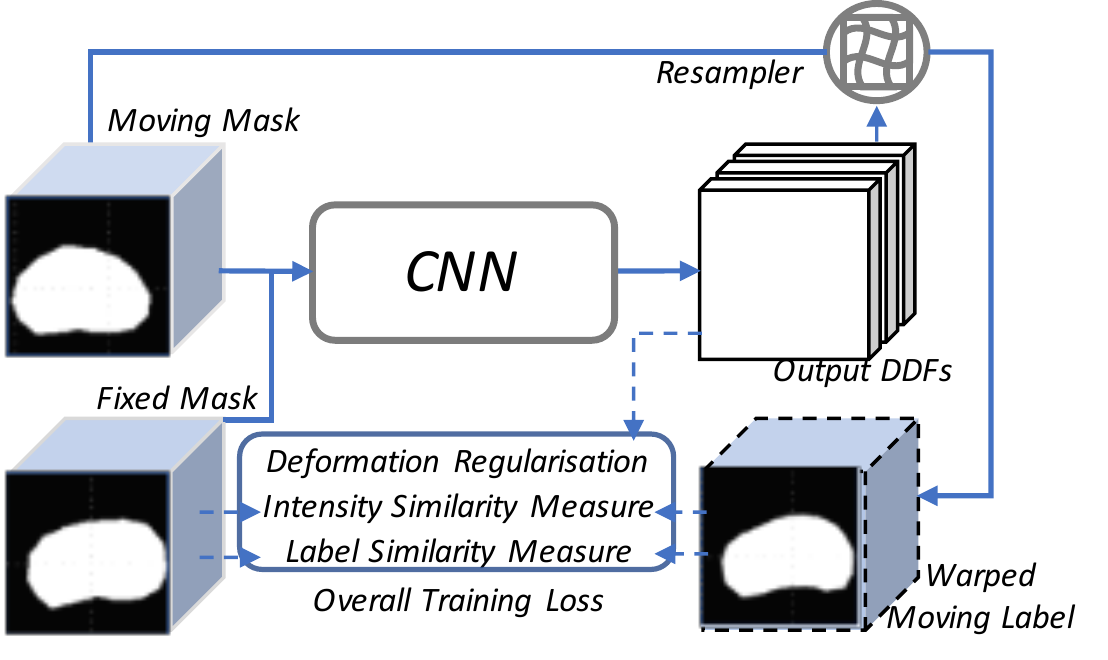}
    \caption{Architecture of the feature-based registration network described in Sec.~\ref{sec:feature-based registration}}
    \label{fig:network}
\end{figure}

In this work, we aim to use KDE for forecasting future morphological changes, which requires registering a large number of image masks to quantify changes in morphology, described in Section 2.2. Therefore, we first train a deep-learning-based pairwise registration network, highly efficient in registering image masks that delineates prostate gland capsules, the image feature of interest. As illustrated in Fig.~\ref{fig:network}, denote $\{(m^{A}, m^{B})\}$ as a set of paired image masks, which are fed into a convolutional neural network during training. The network in turn generates dense displacement fields (DDFs) $\mu^{A \rightarrow B}$, warping moving image $m^{A}$ to the coordinate space of $m^{B}$. The training loss comprises multi-scale Dice and sum-of-squared distance (SSD), to measure mask similarity, with bending energy as deformation regularizer, which are empirically weighted by 1.0, 1.0, and 5.0 based on previous work \cite{yang2020longitudinal}, respectively.

\subsection{Kernel density estimation for DDF forecasting}
In this section, we describe a KDE method \cite{parzen1962estimation} to model the \textit{future morphological changes} in the DDF space. Given a data set $\{(T^i_j, m^i_j), i\in[1,n], j\in[1, O^i]\}$, where $m^i_j$ represents $j^{th}$ image mask from $i^{th}$ patient, acquired at time $T^i_j$ (here, in months). $n$ and $O^i$ are the number of patients and the number of images from $i^{th}$ patient, respectively. Consecutive image masks from individual $i^{th}$ patients are aligned with the registration network described in Sec. 2.1. The resulting $O^i-1$ DDFs $\mu^{i \rightarrow i}_{j \rightarrow j+1}$ describe the intra-subject morphological changes, one we wish to model, between current time $T^i_j$ and future target time $T^i_{j+1}$.

\begin{figure}[htb]
    \centering
    \includegraphics[width=8.5cm]{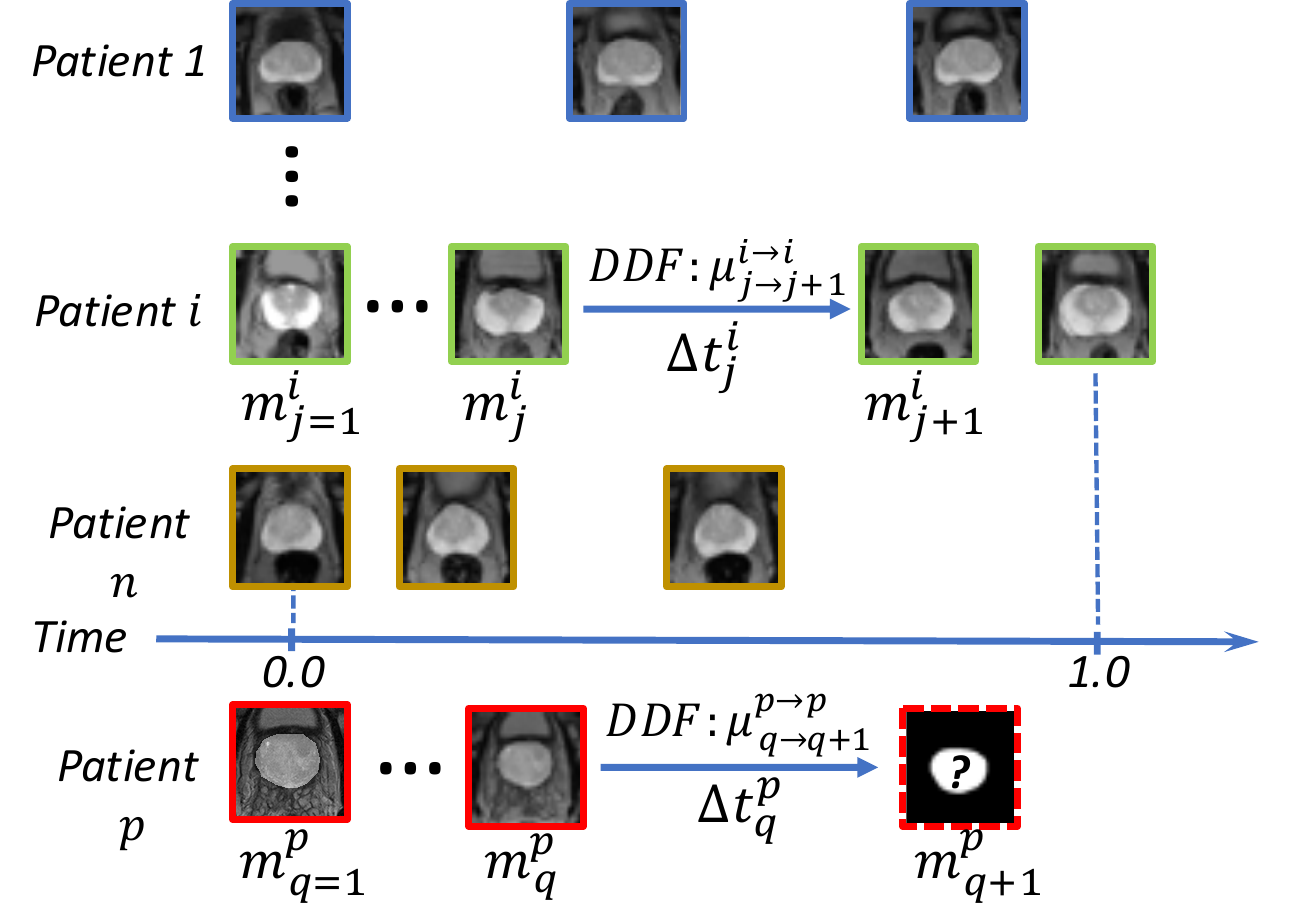}
    \caption{An illustration of the data used in the KDE modelling. Only the gland masks are used during this process, which are annotated from their corresponding MR T2-weighted images.}
    \label{fig:dataset}
\end{figure}

For an image mask $m^p_q$ acquired at time $T^p_q$, without $m^p_{q+1}$ being available, we therefore predict the morphological change towards time $T^p_{q+1}$, by estimating the probability density of the DDF $\mu^{p \rightarrow p}_{q \rightarrow q+1}$ with KDE. Thus, the expected DDF $\bar{\mu}^{p \rightarrow p}_{q \rightarrow q+1}$ is given by Eq.~\ref{eq:kde}:
\begin{equation}
    \bar{\mu}^{p \rightarrow p}_{q \rightarrow q+1}= \sum\limits_{i=1}^{n}\sum\limits_{j=1}^{O^{i}-1}K^{p,i}_{q,j}, \cdot \mu^{i \rightarrow i}_{j \rightarrow j+1}
\label{eq:kde}
\end{equation}
where the normalised kernel $K^{p,i}_{q,j}=k^{p,i}_{q,j}/\sum\limits_{i=1}^{n}\sum\limits_{j=1}^{O^{i}-1}k^{p,i}_{q,j}$ takes a Gaussian form in Eq.~\ref{eq:gk}:
\begin{equation}
    k^{p,i}_{q,j}=\frac{1}{\sigma\sqrt{2\pi}}\exp({-{(D^{p,i}_{i,j})}^2/{2\sigma^2}})
\label{eq:gk}
\end{equation}
Computing the kernel requires a distance function $D^{p,i}_{q,j}$ between data pairs, $(T^i_j, m^i_j)$ and $(T^p_q, m^p_q)$, with a kernel bandwidth parameter $\sigma^2$. As illustrated in Fig.~\ref{fig:dataset}, we propose a novel kernel distance function, which takes into account differences in morphology, stage-of-progression, duration-of-change and whether they are from the same subject.
First, the \textit{morphology} is represented by the image masks $m^i_j$. The DDFs $\mu^{i \rightarrow p}_{j \rightarrow q}$ are considered as a measure of difference in morphology and are computed by the same registration network described in Sec. 2.1. Obtaining these DDFs require both intra-subject (i.e. for available data at previous time points, $i=p$, $j<q$) and inter-subject (i.e. $i \neq p$) registration. Second, the absolute time $T^i_j$ is normalised to $t^i_j=(T^i_j-T^i_1)/(T^x_{O^{x}}-T^x_1)$, where $x$ denotes the patient that has the longest time interval from baseline in the entire data set. Hence, $t^i_j \in [0,1]$ may represent \textit{stage-of-progression}. This normalisation may reduce uninformative variance in difference of baseline time $T^i_{j=1}$, assuming a consistent inclusion criterion for active surveillance program. Third, $\Delta{t^i_j}$ denotes the time interval between $t^i_j$ and $t^i_{j+1}$, representing the \textit{duration-of-change} between two time points. 

Assuming the future morphological changes depend on these factors, which are further assumed to follow independent distributions from exponential family, the proposed kernel distance function is given by Eq.~\ref{eq:distance}:
\begin{equation}
\begin{aligned}
    {(D^{p,i}_{i,j})}^2 = &\alpha\cdot\frac{1}{V}\cdot||\mu^{i \rightarrow p}_{j \rightarrow q}||^2_2\\
    + & \beta\cdot(t^p_q-t^i_j)^2\\
    + & \gamma\cdot(\Delta{t^p_q}-\Delta{t^i_j})^2\\ 
    + & \lambda\cdot(1-0^{|i-p|})
\end{aligned}
\label{eq:distance}
\end{equation}
where, the last term $\lambda$ is a ``subject-specificity bias" hyper-parameter, discriminating whether the pair of data $(T^i_j, m^i_j)$ and $(T^p_q, m^p_q)$ are acquired from the same patient. A positive value is used for inter-subject pairs, to reflect the fact that the morphological changes ought to be more strongly correlated between data from the same patients, zero otherwise. $V$ is the number of voxels of $\mu$. Hyper-parameters $\alpha$, $\beta$, $\gamma$ and $\lambda$ are not fully independent and were experimentally optimised during the model development described in Sec. 3.


\section{Experiment}
\label{sec:experiment}

A total of 331 longitudinal prostate T2-weighted MR images were acquired and manually segmented from 73 patients at University College London Hospitals NHS Foundation. For each patient, 3-9 image masks were available. Intervals between consecutive visits range from 6.6 to 51.3 months, with a mean of 18.05 and a standard deviation of 10.25 months. For computational consideration, all 3D image masks were resampled to $0.7\times0.7\times0.7$ $mm^3$ isotropic voxels and an image size of $128\times128\times102$, after pre-aligning their centres of mass, with an intensity range of $[0,1]$ as soft labels following a trilinear interpolation. The data were split into 53, 9 and 11 patients for training, validation and holdout sets, respectively.

The registration network training was adapted from the DeepReg package \cite{Fu2020Deepreg}, based on TensorFlow 2. A previously proposed multi-scale encoder-decoder architecture was used \cite{hu2018weakly}. The Adam optimizer with an initial learning rate of $10^{-5}$ was used. The network was trained on Nvidia Tesla V100 GPUs with a mini-batch of 4, half of which were sampled from intra-patient pairs with time-forward order, the other half were sampled from inter-patient pairs. The network was trained for 800 iterations, approximately 7.5 hours. The hyper-parameters in Eq.~\ref{eq:distance} were tuned on the validation set using a grid search. $\alpha$, $\beta$, $\gamma$, $\lambda$ and $\sigma$ were set to 1.0, 10.0, 10.0, 0.1 and 1.0, respectively, via gird search. 

The feature-based registration is validated based on Dice similarity coefficient (DSC), between the fixed mask and the registration-warped moving mask, on the holdout set. Quantitative metrics were calculated to demonstrate the efficacy of the KDE-predicted DDFs, on the hold out patients. The relative gland volume error (rVE) is computed between the ground-truth image masks and the KDE-predicted-DDF-warped image masks from previous time points. The predicted volumes are plotted with time, and are compared with those from the ground-truth image masks and those from registration (also requires the ground-truth image masks). The relative volume change (rVC) can be computed between the two time points, defined as zero without prediction or registration. The gland volume is an important clinical measure for prostate cancer progression, such as in calculating prostate-specific antigen density. The Jacobian determinant map was computed to measure the degree of volume change in the predicted DDFs. We also provide two qualitative case studies, for visualizing the predicted morphology changes over baselines and five follow-up visits.

\section{Results}
\label{sec:results}
\textbf{Registration results} Compared to the centroid-aligned image masks without any prediction or registration, with a mean DSC of 0.854$\pm$0.03 and a mean rVE of 9.176$\pm$7.42\%, the proposed registration method achieved improvement in terms of both Dice score (0.920$\pm$0.008) and rVE (2.009$\pm$0.486\%) with statistical significance (paired t-tests, both p-values$<$ 0.001). The trained network registered a pair of 3D masks in approximately 0.2-0.25 second on the same GPUs. 

\textbf{KDE forecasting results} It is noteworthy that the registration results can not be fairly compared with that from the KDE model, since the KDE model predicts the DDFs without the access to image masks acquired at the future time points. We report a mean Dice score of 0.865$\pm$0.028 using the proposed method, improving the centroid alignment with statistical significance (p-value$<$0.001), also with visible improvement on the gland morphological changes, as illustrated in Fig 4. An arguably more evident improvement can be visualised on the gland volume prediction with the KDE; lacking significant improvement in rVE is probably due to the large variance in prostate sizes, affected by many clinical and non-clinical factors such as disease status, force exerted from surrounding organs and treatment interventions. As plotted in Fig.3, the predicted trend of the volume from KDE model is visibly consistent with the ground-truth over all test subjects, with a significantly improved rVC of 3.425$\pm$1.591\% compared with no-prediction (p$<$0.001). 

We also report that, during analysis of the results, one case was identified as an outlier, due to the fact that the patient had indeed received prostatectomy between the second and third visits. Including this case, the mean results from KDE model were 0.859$\pm$0.049, 11.692$\pm$20.259\%, 3.405$\pm$1.554\% and 1.031$\pm$0.003, for DSC, rVE, rVC and difference in Jacobian determinant values, respectively. Fig.~\ref{fig:vis} shows  example  slices  with corresponding Jacobian determinant maps. The registration and KDE model led to Jacobian determinant values of 1.036$\pm$0.014 and 1.031$\pm$0.003 of the  from their corresponding DDFs. No values smaller than zero were found in all cases, consistent with the registration-estimated DDFs.
\begin{figure}[htb]
    \centering
    \includegraphics[width=8.5cm]{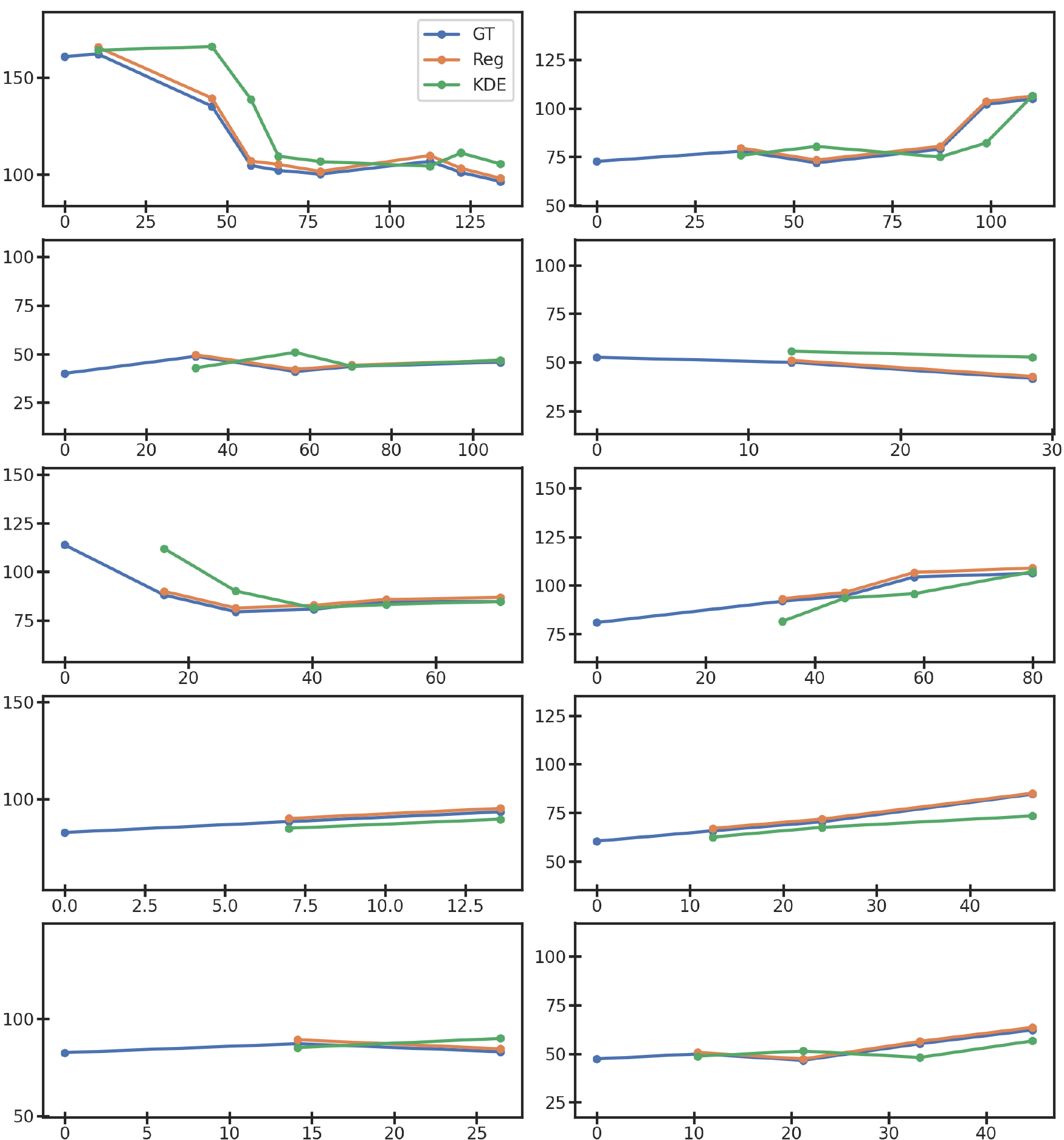}
    \caption{The gland volumes (in $cm^{3}$) calculated from the predictions are plotted over available time points (in months), for patients in the holdout set.}
    \label{fig:size}
\end{figure} 
\section{Conclusion}
\label{sec:conclusion}
In this paper, we propose to forecast future changes in prostate gland morphology, using an efficient feature-based registration and a novel KDE to predict DDFs without data from target time points. With experiment results obtained from a longitudinal active surveillance data set, we propose to generalise the classical KDE algorithm to high-dimensional DDF space, and demonstrate its interesting future-prediction ability on the holdout test set. Future studies aim to further improve the morphology forecasting and as well as to apply morphological forecasting to a wider range of clinical applications.
\begin{figure}[ht!]
    \centering
    \includegraphics[width=8.5cm]{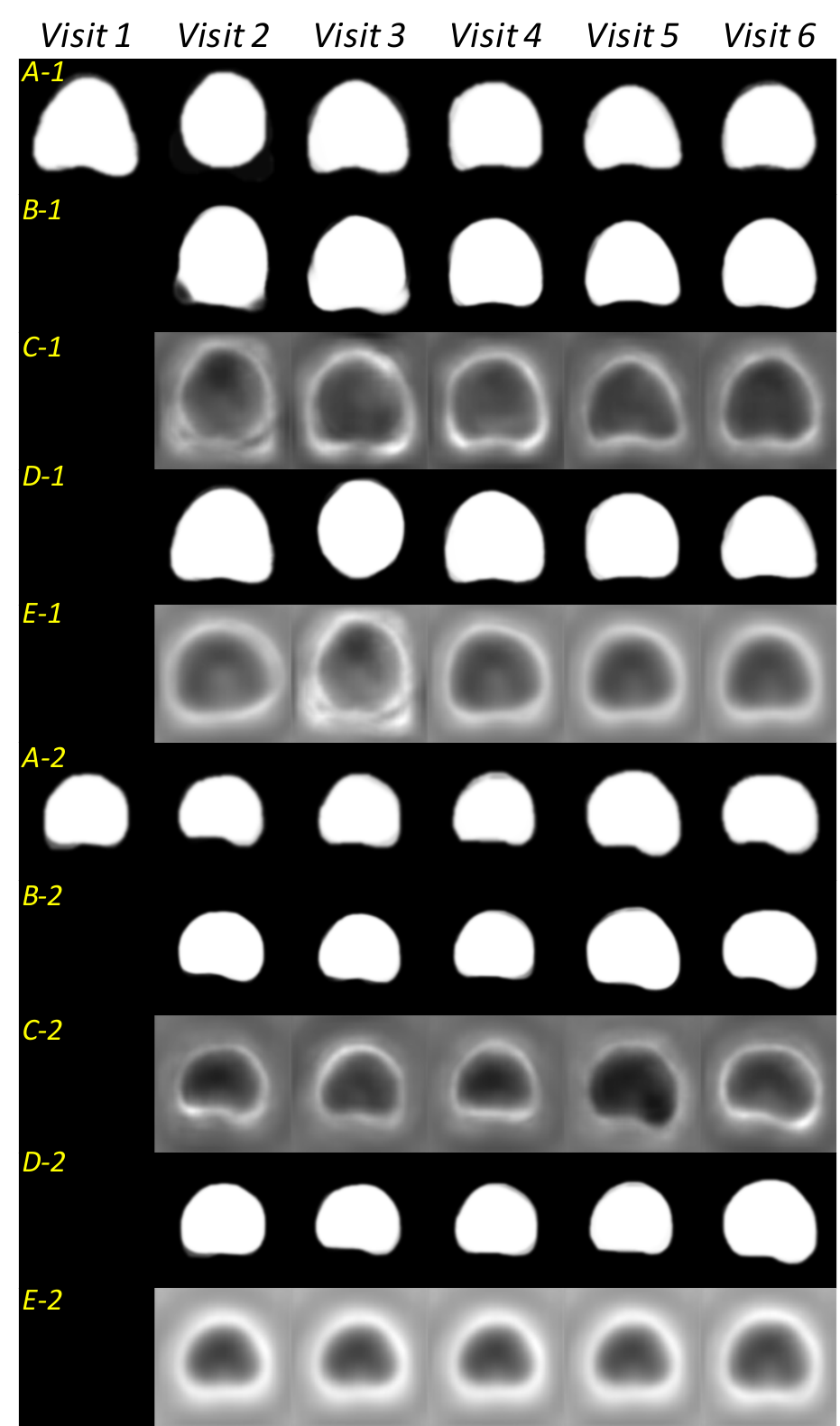}
    \caption{Two example cases over multiple visits in the holdout set. Rows show image masks from ground truth (A), registration (B) and the KED models (D). The latter two are followed by their corresponding Jacobian determinant maps (C) and (E), respectively.}
    \label{fig:vis}
\end{figure} 

\section{Compliance with Ethical Standards}
\label{sec:ethics}
At our institution, all clinical records and MR images are routinely reviewed as part of an audit performed for the internal evaluation of the active surveillance service and no institutional review board approval was required.
\section{Acknowledgments}
\label{sec:acknowledgments}
This work is supported by the Wellcome/EPSRC Centre for Interventional and Surgical Sciences (203145Z/16/Z), Centre for Medical Engineering (203148/Z/16/Z; NS/A000049/1), the EPSRC-funded UCL Centre for Doctoral Training in Intelligent, Integrated Imaging in Healthcare (EP/S021930/1), the Department of Health’s NIHR-funded Biomedical Research Centre at UCLH, and CRUK (C28070/A30912; \\ C73666/A31378). Francesco Giganti is funded by the UCL Graduate Research Scholarship and the Brahm Ph.D. scholarship in memory of Chris Adams.

\bibliographystyle{IEEEbib}
\bibliography{refs}

\begin{thebibliography}{10}

\bibitem{lovegrove2018prostate}
Catherine~Elizabeth Lovegrove, Mudit Matanhelia, Jagpal Randeva, David
  Eldred-Evans, Henry Tam, Saiful Miah, Mathias Winkler, Hashim~U Ahmed, and
  Taimur~T Shah,
\newblock ``Prostate imaging features that indicate benign or malignant
  pathology on biopsy,''
\newblock {\em Translational andrology and urology}, vol. 7, no. Suppl 4, pp.
  S420, 2018.

\bibitem{stavrinides2019mri}
Vasilis Stavrinides, Francesco Giganti, Mark Emberton, and Caroline~M Moore,
\newblock ``Mri in active surveillance: a critical review,''
\newblock {\em Prostate Cancer and Prostatic Diseases}, vol. 22, no. 1, pp.
  5--15, 2019.

\bibitem{yang2020longitudinal}
Qianye Yang, Yunguan Fu, Francesco Giganti, Nooshin Ghavami, Qingchao Chen,
  J~Alison Noble, Tom Vercauteren, Dean Barratt, and Yipeng Hu,
\newblock ``Longitudinal image registration with temporal-order and
  subject-specificity discrimination,''
\newblock in {\em MICCAI}. Springer, 2020, pp. 243--252.

\bibitem{shen2002hammer}
Dinggang Shen and Christos Davatzikos,
\newblock ``Hammer: hierarchical attribute matching mechanism for elastic
  registration,''
\newblock {\em IEEE transactions on medical imaging}, vol. 21, no. 11, pp.
  1421--1439, 2002.

\bibitem{baum2020multimodality}
Zachary~MC Baum, Yipeng Hu, and Dean~C Barratt,
\newblock ``Multimodality biomedical image registration using free point
  transformer networks,''
\newblock in {\em MICCAI 2020 ASMUS Workshop}, pp. 116--125. Springer, 2020.

\bibitem{gu2013contour}
Xuejun Gu, Bin Dong, Jing Wang, John Yordy, Loren Mell, Xun Jia, and Steve~B
  Jiang,
\newblock ``A contour-guided deformable image registration algorithm for
  adaptive radiotherapy,''
\newblock {\em Physics in Medicine \& Biology}, vol. 58, no. 6, pp. 1889, 2013.

\bibitem{mazumder2020validnet}
Joy Mazumder, Mohsen Zand, Sheikh Ziauddin, and Michael Greenspan,
\newblock ``Validnet: A deep learning network for validation of surface
  registration.,''
\newblock in {\em VISIGRAPP (4: VISAPP)}, 2020, pp. 389--397.

\bibitem{hu2018weakly}
Yipeng Hu, Marc Modat, Eli Gibson, Wenqi Li, Nooshin Ghavami, Ester Bonmati,
  Guotai Wang, Steven Bandula, Caroline~M Moore, Mark Emberton, et~al.,
\newblock ``Weakly-supervised convolutional neural networks for multimodal
  image registration,''
\newblock {\em Medical image analysis}, vol. 49, pp. 1--13, 2018.

\bibitem{balakrishnan2019voxelmorph}
Guha Balakrishnan, Amy Zhao, Mert~R Sabuncu, John Guttag, and Adrian~V Dalca,
\newblock ``Voxelmorph: a learning framework for deformable medical image
  registration,''
\newblock {\em IEEE transactions on medical imaging}, vol. 38, no. 8, pp.
  1788--1800, 2019.

\bibitem{orczyk20173d}
Cl{\'e}ment Orczyk, Andrew~B Rosenkrantz, Artem Mikheev, Arnauld Villers,
  Myriam Bernaudin, Samir~S Taneja, Samuel Valable, and Henry Rusinek,
\newblock ``3d registration of mpmri for assessment of prostate cancer focal
  therapy,''
\newblock {\em Academic radiology}, vol. 24, no. 12, pp. 1544--1555, 2017.

\bibitem{vuollo2016analyzing}
Ville Vuollo, Lasse Holmstr{\"o}m, Henri Aarnivala, Virpi Harila, Tuomo
  Heikkinen, Pertti Pirttiniemi, and Arja~Marita Valkama,
\newblock ``Analyzing infant head flatness and asymmetry using kernel density
  estimation of directional surface data from a craniofacial 3d model,''
\newblock {\em Statistics in Medicine}, vol. 35, no. 26, pp. 4891--4904, 2016.

\bibitem{hu2015population}
Yipeng Hu, Eli Gibson, Hashim~Uddin Ahmed, Caroline~M Moore, Mark Emberton, and
  Dean~C Barratt,
\newblock ``Population-based prediction of subject-specific prostate
  deformation for mr-to-ultrasound image registration,''
\newblock {\em Medical image analysis}, vol. 26, no. 1, pp. 332--344, 2015.

\bibitem{parzen1962estimation}
Emanuel Parzen,
\newblock ``On estimation of a probability density function and mode,''
\newblock {\em The annals of mathematical statistics}, vol. 33, no. 3, pp.
  1065--1076, 1962.

\bibitem{Fu2020Deepreg}
Y~Fu, N~Montana~Brown, SU~Saeed, A~Casamitjana, ZMC Baum, R~Delaunay, Q~Yang,
  A~Grimwood, Z~Min, E~Bonmati, et~al.,
\newblock ``Deepreg: a deep learning toolkit for medical image registration,''
\newblock {\em The Journal of Open Source Software}, 2020.

\end{thebibliography}
\end{document}